# Study of the $^{15}$N(p,n)$^{15}$O reaction as a monoenergetic neutron source for the measurement of differential scattering cross sections

Erik Poenitz,[a] Ralf Nolte,[a,*] Dankwart Schmidt,[a] Guochang Chen[b]

[a] *Physikalisch-Technische Bundesanstalt,*
*Bundesallee 100, 38116 Braunschweig, Germany,*
[b] *China Institute of Atomic Energy,*
*Beijing 102413, China*
*E-mail*: Ralf.Nolte@ptb.de

ABSTRACT: The $^{15}$N(p,n) reaction is a promising candidate for the production of monoenergetic neutrons with energies of up to 5.7 MeV at the facilities where the T(p,n)$^3$He reaction cannot be used. The characteristic properties of this reaction were studied focusing on its suitability as a source of monoenergetic neutrons for the measurement of differential scattering cross sections in the neutron energy range of 2 MeV to 5 MeV. For this purpose differential and integral cross sections were measured and the choice of optimum target conditions was investigated. The reaction has already been used successfully to measure of elastic and inelastic neutron scattering cross sections for $^{nat}$Pb in the energy range from 2 MeV to 4 MeV and for $^{209}$Bi and $^{181}$Ta at 4 MeV.

KEYWORDS: $^{15}$N(p,n)$^{15}$O reaction, monoenergetic neutrons, neutron scattering cross sections.

---

[*] Corresponding author.

**Contents**

## 1. Introduction

The Physikalisch-Technische Bundesanstalt (PTB) operates a time-of-flight (TOF) spectrometer for the measurement of elastic and inelastic neutron scattering cross sections and double-differential neutron emission cross sections. In the past 30 years, cross sections were measured for a variety of elements [1, 2] using the D(d,n)$^3$He reaction as a source for quasi-monoenergetic neutrons. The measurements were carried out with a focus on the energy region between 8 MeV to 14 MeV where data are scarce due to the unavailability of a monoenergetic neutron source. Because of the deuteron energy range available at the PTB cyclotron and the $Q$ value of the reaction, $Q$ = +3.27 MeV, neutrons can be produced in the energy range from 6 MeV to 16 MeV.

So far, the measured elements were selected with a focus on materials of interest for fusion technology, i.e. Fe, W, Be and many more. The experimental cross section data measured at the PTB TOF spectrometer exhibit relatively small uncertainties compared with data sets measured at other facilities. This is primarily due to the well characterized properties of the entire spectrometer comprising a realistic description of the production, interactions and detection of the neutrons.

The aim of the present work is the extension of the available neutron energy range to lower energies by introducing a new neutron source which facilitates the production of neutrons with energies between 2 MeV and 5 MeV using the ion beams available from the PTB cyclotron. The extension of the energy range enables the measurement of scattering cross sections for $^{nat}$Pb, $^{209}$Bi and $^{181}$Ta in the energy range from 2 MeV to 5 MeV. Reliable cross section data in this energy range are required for the design of lead-cooled fast reactors (LFR) and accelerator-driven subcritical reactors (ADS). In particular, the neutron transport in the lead spallation target of an ADS strongly depends on the inelastic neutron scattering cross sections in the energy region from 0.5 MeV to 6 MeV [3, 4].

## 2. Experimental procedure

### 2.1 The PTB TOF spectrometer

The PTB TOF spectrometer is optimized for the measurement of neutron scattering cross sections in the energy range from 6 MeV to 16 MeV. An overview of the TOF spectrometer is shown in figure 1. A more deailed description of the TOF spectrometer can be found in [1].

The cyclotron (CY) can produce proton (p), deuteron (d) and α-particle beams (α) in the energy ranges $E_p$ = 2 MeV to 19 MeV, $E_d$ = 3 MeV to 13.5 MeV and $E_\alpha$ = 6 MeV to 26 MeV, respectively. The repetition frequency of the beams can be adapted to the needs of TOF measurements using an internal beam deflector. With a repetition frequency of about 1 MHz and a flight path of 12 m, time frame overlap can be avoided for neutron energies above 700 keV.

The neutrons are produced in a gas target (T), 30 mm in length. A molybdenum entrance foil 5 μm in thickness is used to separate the gas cell from the accelerator vacuum. After passing the gas cell, the beam is stopped in a gold foil, 0.5 mm in thickness. The compact construction of the cyclotron prevents the installation of an analysis magnet in the beam line. Hence, the precise energy of the incident proton beam has to be calculated from the measured neutron energy at an emission angle of 0° and the energy loss of the protons in the entrance foil and the target gas.

The neutron detectors D1-D5 are located at a distance of 12 m from the pivot (S) of the spectrometer. The angular separation of the detectors is 12.5°. They are shielded against the source by a massive collimator system consisting of a polyethylene pre-collimator (P), tanks with borated water (W) and a concrete wall (CO). The scattering or neutron emission angle is varied by rotating the cyclotron on a moveable arm.

The detector D1 to D5 as well as the monitor detector (M) are NE213 liquid scintillation detectors which are sensitive to photons and fast neutrons. Pulse-shape discrimination techniques are used to distinguish between photon and neutron induced events. Detector D1 has a diameter of 10.16 cm and a thickness of 2.54 cm. In the measurements of scattering cross sections using the D(d,n)$^3$He reaction as a neutron source, it is used for measurements of neutrons directly emitted from the target and for the measurement of differential scattering cross sections at forward angles where the elastic scattering cross sections are sufficiently high. Detectors D2 – D5 are larger and have a diameter of 25.4 cm and a thickness of 5.08 cm. The monitor detector has a diameter of 3.81 cm and a thickness of 3.81 cm. It is shielded by a polyethylene collimator and pre-collimator and is independently moveable. In organic scintillators, the frame overlap energy of about 700 keV corresponds to electron-equivalent energy of about 100 keV for recoil protons. This is well below the pulse-height threshold of 170 keV which was used in the present work for the detectors D1 – D5.

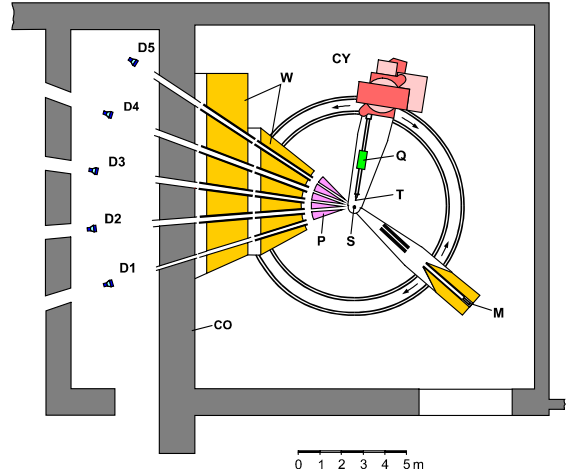

**Fig. 1** Schematic view of the accelerator-based neutron source and the time-of-flight spectrometer arrangement for the measurement of differential scattering cross sections. The distance between the pivot axis S and the geometrical centers of the sensitive volumes of the detectors D1 – D5 is 12 m. The angular separation between the detectors is 12.5°. The quadrupole magnet Q is used to transport the proton beam from the CV28 cyclotron to the gas target T. The detector M is used to monitor the emission rate of the source. The collimators consist of polyethylene slabs (P) and water tanks (W) with steel inserts.

## 2.2 Properties of the $^{15}$N(p,n)$^{15}$O reaction as a neutron source

The most important requirement for the new neutron source was a large negative $Q$-value and sufficient neutron yield at forward direction. Among the neutron production reactions included in the DROSG-2000 code [5], the $^{15}$N(p,n)$^{15}$O reaction comes closest to these requirements. The (d,n) and (α,n) reactions are usually ruled out because of their positive $Q$-values and small neutron yield with the alpha beam currents available at the PTB cyclotron, respectively.

The $Q$-value of the $^{15}$N(p,n) reaction is -3.54 MeV. Hence, with the 2 MeV to 19 MeV proton beams available at the PTB cyclotron, neutrons with energies $E_n$ from 0.01 MeV to 15.5 MeV can be produced. A production of monoenergetic neutrons is possible for proton energies $E_p \leq 9.31$ MeV ($E_n(0°) \leq 5.74$ MeV). At higher energies, neutron production is also possible via excited states of $^{15}$O. Because of the neutron detection thresholds of the NE213 detectors of approximately 0.7 MeV, the $^{15}$N(p,n)$^{15}$O reaction is practically monoenergetic at even somewhat higher energies. In practice, there is no gap to the neutron energy range available via the D(d,n)$^3$He reaction.

The $^{15}$N(p,n)$^{15}$O reaction also has certain disadvantages compared to the D(d,n)$^3$He reaction. For example, the electronic stopping power of protons in nitrogen gas is larger than the one of deuterons in deuterium gas [6] which leads to a reduced neutron yield. The stopping power for deuterons in D$_2$ gas and protons in $^{15}$N$_2$ gas in the interesting energy ranges is depicted in figure 2. Both calculations were carried out for a gas pressure of 1000 hPa at 20 °C.

For the measurement of differential scattering cross sections using the D(d,n)$^3$He source, a gas pressure of 2000 hPa is used. This corresponds to an energy loss in the gas cell ranging from 120 keV for $E_n(0°) = 6$ MeV to 40 keV for $E_n(0°) = 14$ MeV. The neutron energy width is also influenced by other factors, for example the width of the energy distribution of the projectiles, the energy broadening caused by reaction kinematics and the angular interval covered by the scattering sample. The energy width is approximately 110 keV to 140 keV [7, 8] and is dominated by the energy loss in the gas target for low energies and by the reaction kinematics of the D(d,n)$^3$He reaction for high energies.

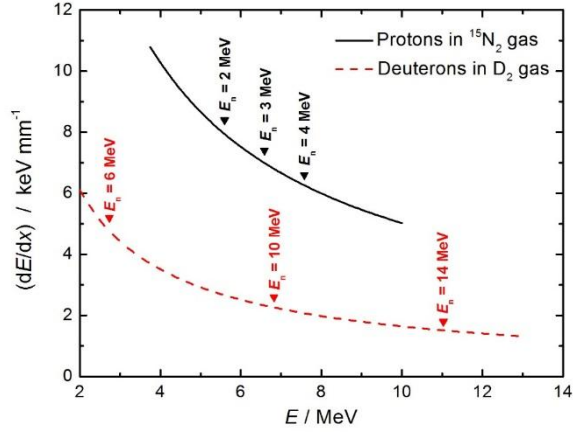

**Fig. 2** Differential energy loss (d$E$/d$x$) of protons in $^{15}N_2$ (black solid curve) and deuterons in $D_2$ (red dashed curve) as a function of proton or deuteron energy $E$. Both curves were calculated for a gas pressure of 1000 hPa. For a better comparison, 0° neutron energies are indicated for both neutron producing reactions.

For measurements of differential inelastic scattering cross sections for $^{206,207}$Pb [9, 10] using the $^{15}$N(p,n)$^{15}$O source, a separation of the first excited level of $^{206}$Pb ($E_x$ = 803 keV) and the second excited level of $^{207}$Pb ($E_x$ = 898 keV) has to be achieved. Therefore, the energy width had to be reduced to approximately 90 keV. This requires a reduced gas pressure of 400 hPa to 500 hPa. In this case, the neutron energy width is determined by the energy loss in the gas target for all projectile energies in the energy range of interest.

The cross section of the D(d,n)$^3$He reaction is smooth and only weakly energy dependent. The angular distribution is always forward-peaked. In contrast, the cross section of the $^{15}$N(p,n)$^{15}$O reaction shows many narrow resonances and the angular distributions are not always forward-peaked. Therefore, measurements of differential scattering cross section can only be carried out at selected energies where the differential cross section at 0° is sufficiently large. Even at these energies, the neutron yield at 0° is an order of magnitude lower than for the D(d,n)$^3$He reaction. Therefore, longer measurement times than with the D(d,n) source are required.

The procedure used for the analysis of measurements of differential scattering cross sections at the PTB TOF spectrometer includes the complete simulation of the experiment using the Monte Carlo code SINENA [11] and the iterative adjustment of the scattering cross sections for the element under study. For this purpose, SINENA uses cross section data sets for the simulation of the neutron production in the gas target. These had to be extended for the implementation of the new neutron producing reaction. Hence, the consistency of the existing cross section data is an important issue. Experimental data for $^{15}$N(p,n)$^{15}$O cross sections are available [12 – 19], but none of them were measured with the focus on the use of the reaction as a neutron source. For the measurements reported in [12, 13, 15, 19], thin targets ($\Delta E$ = 10 keV to 40 keV) were used while for the other measurements thicker targets ($\Delta E \approx$ 130 keV to 480 keV) were employed which resulted in averaging over the resonances. In [15, 19], 511 keV annihilation photons resulting from the $\beta^+$ decay of $^{15}$O were measured. These activation measurements provided cross section data but no information about angular distributions. For proton energies $E_p \geq$ 9.31 MeV, production of $^{15}$O via excited states and subsequent de-excitation to the ground state is possible. In this case, the production cross section measured

– 4 –

using the activation technique is larger than the partial neutron emission cross section measured by detection of the neutrons produced in transitions to the ground state of $^{15}$O. The uncertainties range from 50 % for the measurements by Jones *et al.* [12] (published in 1958) to 3 % to 6 % for the ones by Byrd *et al.* [18] (published in 1981).

In figure 3 differential cross sections from the data compilation of the code DROSG-2000 [5] and from Jones *et al.* [12] are depicted for comparison. The 0° excitation function by Jones *et al.* consists of 215 data points in the energy range $E_p$ = 3.7 MeV to 6.4 MeV ($E_n(0°)$ = 0 MeV – 2.78 MeV) and shows the resonances in the cross sections. The data taken from DROSG-2000 include 85 data points mainly from [12, 14, 16, 17, 18]. The interpolation is done by a spline-fit. The figure demonstrates that the data are not always in agreement, i.e. around $E_p$ = 5.8 MeV ($E_n(0°)$ = 2.2 MeV). Because of the discrepancies in the existing cross section data, measurements of differential cross sections were carried out for selected energies in the energy range $E_p$ = 5.6 MeV to 9.1 MeV. Another goal of the measurements was to identify possible problems related to use this reaction as a neutron source in scattering experiments.

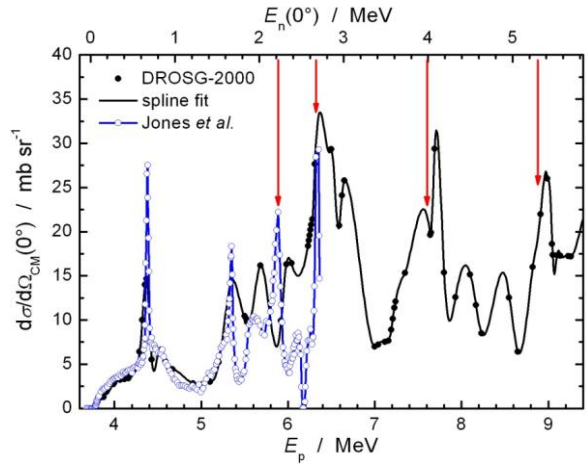

**Fig. 3** Differential cross sections for 0° for the $^{15}$N(p,n) reaction taken from the data base of the code DROSG-2000 (black solid circles) and the measurement by Jones *et al.* (blue circles). The red arrows mark energies with a particularly large differential neutron yield at 0°. These energies are best-suited for the use of this reaction as a neutron source.

**2.3 Measurements and data analysis**

The measurements were carried out at the PTB TOF spectrometer with the gas target positioned at the location of the scattering sample. For most of the measurements, cyclotron angles of 0°, 6°, 60° and 110° were used. Therefore, the range of neutron emission angles extended from 0° to 160°. For each energy, measurements were carried out with a filled and an evacuated gas target to subtract background neutrons originating from interaction of protons with the entrance and beam stop foils of the gas target or other parts of the beam line. The monitor detector D6 was usually positioned at an angle of 90° relative to the proton beam and used to normalize measurements at cyclotron angles larger than 0° to those at a cyclotron angle of 0°.

The energy deposition by the protons causes local heating and a decrease of the local density of the gas [20, 8]. To correct for this effect, measurements at a cyclotron angle of 0° were carried out for different target currents. The current was varied by changing the repetition frequency of the proton beam. The results of these measurements were linearly extrapolated to zero beam current and used to derive a correction for the decrease in $^{15}N_2$ gas density caused by beam heating [21].

For each neutron detector, the parameters pulse height (PH), pulse shape (PS) and time-of-flight (TOF) were measured using standard NIM modules for analogue signal processing. The signals from all six detectors were routed to a PC-based multi-parameter data acquisition system using an analogue multiplexer which also provided identification tags for the individual detectors. The stop signal for the TOF measurements was derived from an inductive pickup located in front of the gas target. The integrated beam current was measured using a calibrated current integrator. Scalers were used to measure the rate of valid stop events and the event rates of all detectors for dead time correction.

In figure 4, typical TOF distributions measured using one of the large NE213 detectors are shown for a filled (gas-in) and an evacuated (gas-out) gas target. Both distributions are normalised to the same integrated target current. Photon-induced events were suppressed by analogue pulse shape discrimination. The broad bump visible in both TOF distribution between TOF bins 0 and 580 are caused by neutron production in (p,n) reactions in the beam stop (0.5 mm thick gold foil at the end of the gas target) and the molybdenum entrance foil of the gas target. The peak at TOF bin 865 results from photons produced in (p,p′γ) and (p,γ) reactions in the gas target and is visible because of the incomplete suppression of photon-induced events. The flat background above channel 600 is caused by incomplete suppression of photon-induced events as well. This background is larger for the measurement with the filled gas-target because of the contribution from the $e^+e^-$ annihilation photons resulting from the $\beta^+$ decay of $^{15}O$. The peaks at TOF bins 150, 760 and 825 are caused by incomplete suppression of beam pulses by the internal pulse selector of the cyclotron (so-called satellite pulses).

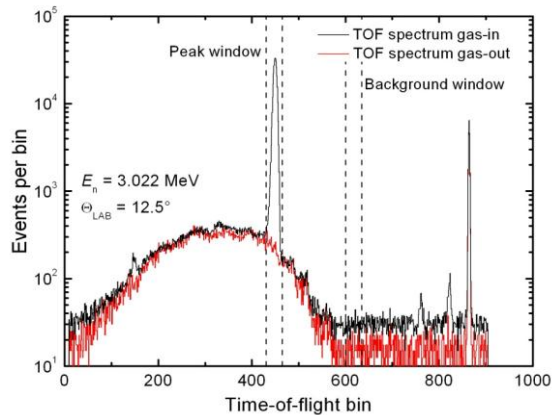

**Fig. 4** Time-of-flight distributions for the large detector D2 with a filled (black histogram) and an evacuated (red histogram) gas cell. The peak and background windows are the time-of-flight regions used for the selection of pulse-height spectra. Photon-induced events were suppressed using an analogue pulse-shape discrimination technique.

The data analysis consisted of the following steps: subtraction of the constant background, dead-time correction (usually smaller than 4 %), correction for unsuppressed proton beam pulses (1 % to 3 %), subtraction of the empty-target background using a measurement with evacuated gas cell and determination of the number of neutrons incident on the detector from the measured data. The last step requires modelling of the neutron detection process in the scintillation detectors using the Monte Carlo method. For the present work the codes NRESP7 and NEFF7 [22] were used. These codes were developed at the PTB and frequently

– 6 –

benchmarked by comparing neutron fluence measurements with organic scintillation detector to measurements with a primary reference instrument, for example a recoil proton telescope. The required input data, i.e. light output functions and PH resolution parameters, were determined carefully for the detectors of the PTB TOF spectrometer. Depending on the neutron energy, two procedures were used to determine the neutron fluence from the measured net PH or TOF distributions.

For the measurements with neutrons of higher energies, PH distributions normalized to one neutron incident on the detector were calculated using NRESP7 and fitted to the experimental net PH distributions. Hence, the scaling factor obtained from this fit yields the number of neutrons incident on the detector. In the present experiment the neutron energies were below or just above the energy of the first excited state in $^{12}$C at 4.439 MeV. For these energies the response of the detectors is entirely determined by elastic scattering on hydrogen and carbon nuclei. The differential cross section for these reactions are known with small uncertainties. Hence, the Monte Carlo simulation describes the pulse-height distribution very accurately. Moreover, this technique does not require a very accurate calibration of the pulse height in electron-equivalent kinetic energy and the fit does not depend strongly on the selection of the fitting region. Therefore, the lower boundary of the fitting region could be set close to the software PH threshold. This technique can be successfully employed if the PH distributions exhibits a flat plateau adjacent to the recoil edge. In such cases, the total relative uncertainty of number of neutron resulting from the analysis procedure is estimated to be about 2 %.

At the lowest neutron energies of the present work, however, the fitting procedure became unreliable because the recoil edge came close the software PH threshold. This is demonstrated in the left panel of figure 5 which shows measured calculated PH distribution for neutron energies of 4.00 MeV and 1.43 MeV. Obviously, the fits for the lower neutron energy depend strongly on the particular weighting procedure and do not reproduce the measured PH distribution. In such cases TOF spectra for a fixed PH detection threshold were used. The effective electron-equivalent energy of the threshold PH was determined using TOF distributions obtained with 'tagged' neutrons from a fast $^{252}$Cf ionisation chamber [1], i.e. relative to the $^{252}$Cf prompt fission neutron spectrum which is considered to be a standard. The uncertainty of the PH threshold determination is estimated to be about 10 keV. The right panel of figure 5 shows detection efficiencies calculated using NEFF7 [22] for PH thresholds corresponding to electron-equivalent energies of 170 keV and 500 keV. The uncertainty band shown for the 170 keV detection threshold was produced by varying the detection threshold by ±10 keV. Depending on how close the recoil edge came to the detection threshold, the uncertainty of the number of neutrons amounted up to 15 % at maximum. In practice, both methods were always used to determine the number of neutrons and the result with the smaller uncertainty was used finally.

The data obtained by the analysis method were corrected for local heating of the target gas (usually 5 % to 14 %), for the isotopic composition ($^{15}$N enrichment greater than 98 %) and chemical purity (impurity concentration less than 0.25 %) of the $^{15}$N$_2$ gas and for outscattering of neutrons in the gas target and in air (12 m flight path). The correction for outscattering of neutrons by the materials of the gas target set up was calculated using the neutron transport code MCNP5. Because of the low-mass target construction, the correction was in the order of 1 % for most of the emission angles. Only for backward angles around 160° the neutrons had to pass more massive parts of the target. Here the correction exceeded 100 %.

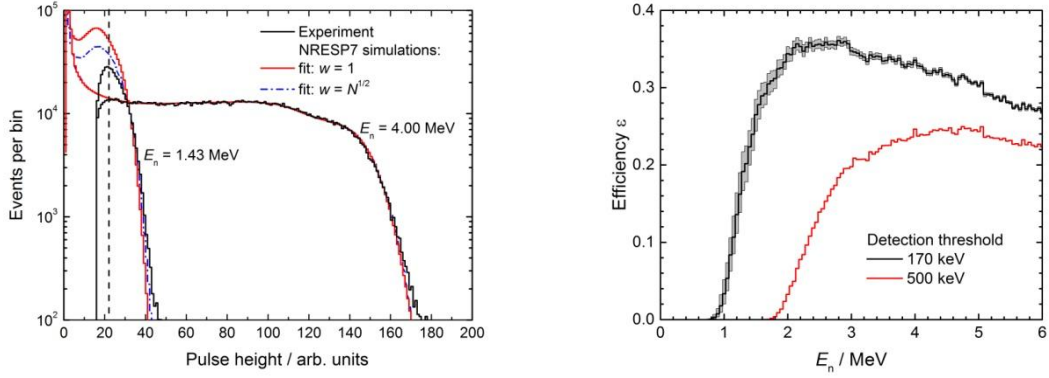

**Fig. 5** The left panel shows measured and calculated pulse height distributions measured with one of the large detectors D2 – D5 for two neutron energies. For the fits constant weights for the experimental data (red solid histograms) and weights proportional to $N^{1/2}$ (blue dashed-dotted histograms) were used. The dashed vertical line indicates the lower boundary of the fitting region. The right panel shows the detection efficiencies for detectors D2 – D5. The pulse height thresholds are 170 keV and 500 keV. Experimentally, the thresholds can be determined with an uncertainty of about 10 keV. The grey band indicates the uncertainty of the detection efficiency for the 170 keV threshold due to the uncertainty of the threshold.

Differential cross sections $(d\sigma/d\Omega_{CM})(\Theta_{CM})$ in the center-of-mass system were calculated from the number of neutrons incident on the detectors, the number $^{15}$N nuclei per unit area, the beam charge and solid angle covered by the neutron detectors. Angle-integrated cross sections were determined by fitting a Legendre polynomial expansion to the data using a least-squares method. Information on the contributions to the total uncertainty of the differential cross sections is provided in table 1.

**Table 1** Uncertainty budget of the measured differential cross sections. The uncertainty contribution are standard measurement uncertainties ($k = 1$), i.e. they correspond to one standard deviation. The uncertainty of the neutron emission angle is 0.2°.

|  | Correction | Contribution to total relative uncertainty |
|---|---|---|
| Number of detected neutrons |  | << 1 % (D1 - D5), 0.5 % - 1.6 % (D6) |
| Unsuppressed proton beam pulses | 1 % - 3 % | 1 % - 3 % |
| Integrated beam current |  | 0.3 % |
| Solid angle of the neutron detectors |  | 0.25 % (D1), 0.1 % (D2 - D5) |
| Local $^{15}$N$_2$ density | 5 % - 14 % | 3 % [1)] |
| Effective length of the gas target (curvature of the entrance foil) | 0.5 mm | 1.6 % |
| Detection efficiency of the neutron detectors |  | 2 % [2)] |
| Outscattering of neutrons | 1 % | 0.1 % [3)] |

[1)] 7 % for the measurements at $E_p$ = 7.57 MeV, 8.7 MeV

[2)] valid for higher neutron energies where PH distribution could be used to determine the neutron fluence. Larger uncertainities of up to 15 % were obtained for the lower neutron energies when TOF distributions for a fixed threshold had to be used.

[3)] larger uncertainties for $\Theta_{LAB} = 160°$ because of reduced transmission through target flange

# 3. Results

## 3.1 General remarks

The measurements of angular distributions were carried out in three measurement campaigns. The results are discussed in the order of increasing projectile energies. The results of the present work are compared with data from the literature, if available. The aim of the measurements was not only to determine the differential neutron emission cross sections of $^{15}$N(p,n)$^{15}$O but also to identify potential problems related to using this reaction as a neutron source for the measurement of differential scattering cross sections. The experimental differential cross sections in the center of mass system (d$\sigma$/d$\Omega_{CM}$)($\Theta_{CM}$), the Legendre coefficients $a_l$, the 0° differential neutron production cross section (d$\sigma$/d$\Omega_{CM}$)(0°) in the center of mass system and the angle-integrated cross sections $\sigma$ are available as an entry [23] in the EXFOR data base maintained by the International Atomic Energy Agency (IAEA).

## 3.2 Proton energies $E_p$ = 5.56 MeV to 5.66 MeV

Measurements were carried out at mean proton energies $E_p$ of 5.56 MeV, 5.62 MeV and 5.66 MeV. These proton energies correspond to neutron energies $E_n$(0°) of 1.96 MeV, 2.03 MeV and 2.06 MeV. Differential cross sections of the $^{15}$N(p,n)$^{15}$O reaction are shown in figure 6. The angular distributions for all three energies exhibit relatively small differential cross sections at $\Theta_{LAB}$ = 0°. Because of the low neutron yield, these energies are not well suited for the measurement of differential scattering cross sections using the $^{15}$N(p,n)$^{15}$O neutron source.

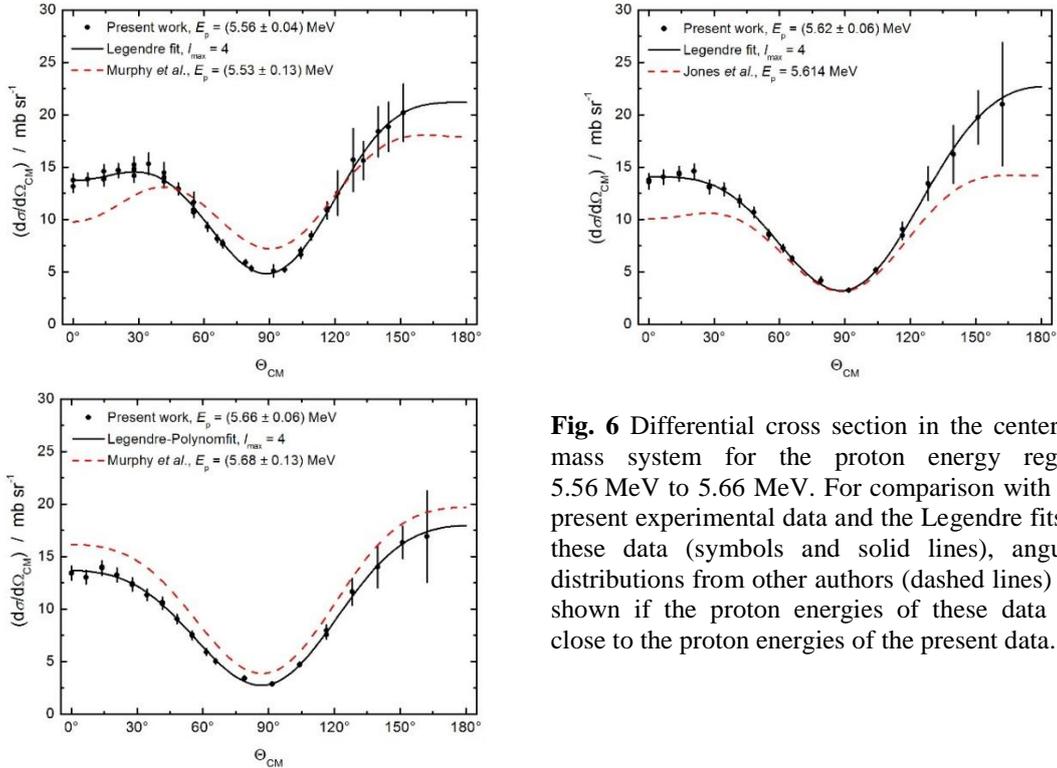

**Fig. 6** Differential cross section in the center of mass system for the proton energy region 5.56 MeV to 5.66 MeV. For comparison with the present experimental data and the Legendre fits to these data (symbols and solid lines), angular distributions from other authors (dashed lines) are shown if the proton energies of these data are close to the proton energies of the present data.

In all three figures, the data points for large angles show relatively large uncertainties. This is caused by the low energy of the neutrons emitted at large angles which are close to the pulse height threshold of the large neutron detectors D2 – D5.

In addition to the data of the present work, angular distributions measured by Murphy *et al.* [14] are shown. The angular distribution reported for a proton energy of 5.53 MeV exhibits a smaller differential cross section for 0° than the one of the present work. This may be caused by the somewhat smaller energy, larger target thickness and the strong energy dependence of the 0° excitation function measured by Jones *et al.* [12] (see figure 3).

The angular distribution reported by Murphy et al. [14] for a proton energy of 5.68 MeV, however, shows a very similar shape of the angular distribution but are 20 % - 25 % larger. The angular distribution reported by Jones et al. [12] for a proton energy of 5.614 MeV is flatter than the one of the present work.

### 3.3 Proton energy $E_p$ = 5.82 MeV

This proton energy corresponds to a neutron energy $E_n(0°)$ = 2.23 MeV. The differential cross sections for $E_p$ = 5.82 MeV are shown in figure 7. The angular distribution is strongly forward-peaked. The 0° cross section is about a factor of two higher than for the measurement at $E_p$ = 5.66 MeV. Because of the strong variation of the angular distribution with projectile energy, the proton energy has to be carefully chosen to achieve a large neutron yield. In addition to the present data, angular distributions by Jones *et al.* [12] and Hansen *et al.* [16] are shown. The angular distribution by Jones *et al.* (stated uncertainty 50 %) is also forward-peaked, but the angular distribution by Hansen *et al.* (uncertainty 10 %) differs in shape and absolute values. This cannot be explained with the different projectile energies and target thicknesses. The measurement by Jones *et al.* is not included in the code DROSG-2000. The measurement by Hansen *et al.* (multiplied with a factor of 4.25) is included in the code and explains the rather small differential cross section at 0° predicted by this code (see solid black line in figure 3).

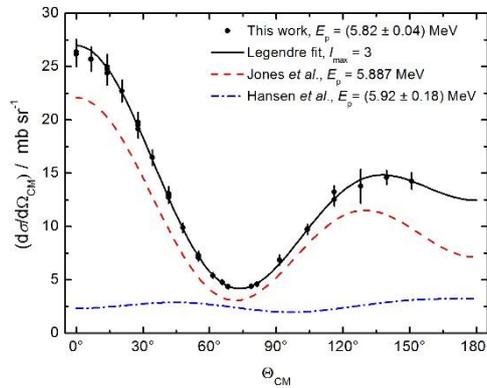

**Fig. 7** Differential cross sections of the $^{15}$N(p,n)$^{15}$O reaction for a proton energy 5.82 MeV. The red dashed line shows the angular distribution measured by Jones et al. for a proton energy 5.887 MeV and the blue dashed-dotted line the one measured by Hansen et al. for a proton energy 5.92 MeV.

### 3.4 Proton energies $E_p$ = 6.62 MeV, 6.69 MeV

These proton energies correspond to neutron energies $E_n(0°)$ of 3.03 MeV and 3.10 MeV. In the left panel of figure 8, differential cross sections for $E_p$ = 6.62 MeV are shown. The angular distribution is forward-peaked but the differential cross section for 0° $(d\sigma/d\Omega_{CM})(0°)$ = 21.7 mb/sr is smaller than for $E_p$ = 6.32 MeV. With exception of the data point for $\Theta_{LAB}$ = 160° the uncertainties of the differential cross sections are in the order of 4 % - 5 %. The large uncertainties for the 160° data points are caused by the correction for the flux attenuation in the flange of the gas target. The data of the present work are compared with an angular distribution measured by Murphy *et al.* [14] for $E_p$ = 6.62 MeV. The angular distribution has a similar shape but the differential cross sections are approximately 15 % larger.

In the right panel of figure 8, the differential cross sections for $E_p$ = 6.69 MeV are shown. The angular distribution is forward-peaked but the differential cross section for 0° $(d\sigma/d\Omega_{CM})(0°)$ = 20.7 mb/sr is smaller than the one for $E_p$ = 6.62 MeV. This is not in agreement with the data in DROSG-2000 which show a minimum at $E_p$ = 6.58 MeV and a maximum at $E_p$ = 6.65 MeV. The present data are compared with an angular distribution measured by Hansen *et al.* [16] for $E_p$ = 6.65 MeV which shows a larger 0° differential cross section. The data points at $E_p$ = 6.62 MeV by Murphy *et al.* and 6.65 MeV by Hansen *et al.* determine the shape of the 0° excitation function in this energy range in the code DROSG-2000 (see figure 3).

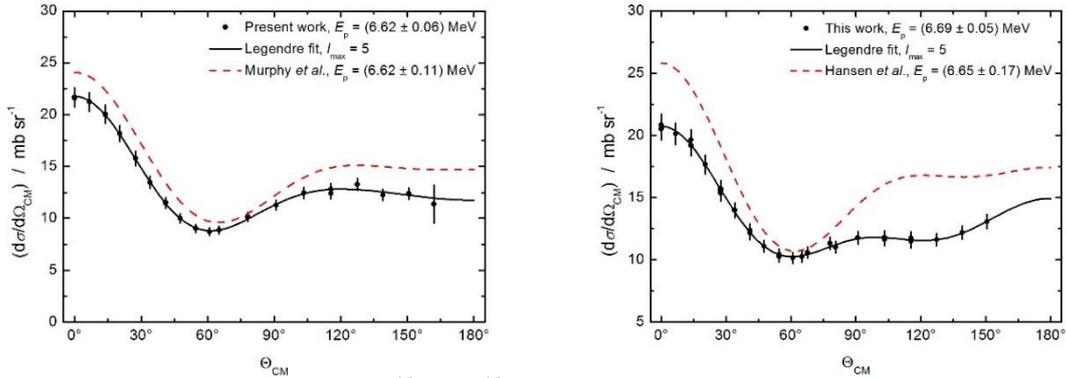

**Fig. 8** Differential cross section of the $^{15}$N(p,n)$^{15}$O reaction for proton energies 6.62 MeV (left panel) and 6.69 MeV (right panel). The red dashed lines show the angular distribution measured by Murphy *et al.* for a proton energy 6.62 MeV (left panel) and by Hansen *et al.* for a proton energy of 6.65 MeV (right panel).

### 3.5 Proton energies $E_p$ = 7.53 MeV to 7.62 MeV

Measurements were carried out for proton energies covering the resonance at $E_p$ = 7.59 MeV. The proton energies $E_p$ = 7.53 MeV, 7.55 MeV, 7.57 MeV, 7.59 MeV, 7.62 MeV and 7.69 MeV correspond to neutron energies $E_n(0°)$ = 3.95 MeV, 3.98 MeV, 3.99 MeV, 4.02 MeV, 4.04 MeV and 4.11 MeV. The results for $E_p$ = 7.53 MeV to 7.62 MeV shown in figure 9 exhibit forward-peaked angular distributions. The largest differential cross section for 0° is reached in the measurement at $E_p$ = 7.57 MeV.

The measurement at $E_p$ = 7.69 MeV is of particular interest. The angular distribution is not forward-peaked. The differential cross section for 0° is rather small, $(d\sigma/d\Omega_{CM})$ = 10.2 mb/sr. This is in contradiction to the data for $E_p$ = 7.69 MeV by Murphy *et al.* [14] which shows a forward-peaked angular distribution. These data determine the predictions of the DROSG-2000 code around this energy. The disagreement is surprising, as the measurements by Murphy *et al.* have small uncertainties (5 % - 7 % for angle-integrated cross section, 4 % for the shape of the angular distribution). The data of this work show a single resonance instead of the double-humped peak predicted by the DROSG-2000 for $E_p \approx$ 7.6 MeV.

### 3.6 Proton energy $E_p$ = 8.70 MeV

Differential cross section for $E_p$ = 8.70 MeV are shown in figure 10. This proton energy corresponds to a neutron energy $E_n(0°)$ of 5.13 MeV. The measured 0° differential cross section is only about 4 mb/sr which is even smaller than the one measured by Wong *et al.* [17] for $E_p$ = 8.65 MeV. This is in clear contradiction to the 0° excitation function in figure 3 which shows a resonance for $E_p \approx 9$ MeV. In test measurements carried out at the PTB TOF spectrometer for $E_p = 8.8$ MeV to 9.1 MeV, differential cross sections for 0° between 18 mb/sr and 22 mb/sr were obtained. In another test measurement at $E_p = 8.51$ MeV, a 0° differential cross section of $(9.1 \pm 0.8)$ mb/sr was determined. Obviously, the excitation function depends more strongly on energy in this energy range than predicted by the DROSG-2000 code. Unfortunately, for the present measurement an energy was chosen which coincides with a minimum in the excitation function.

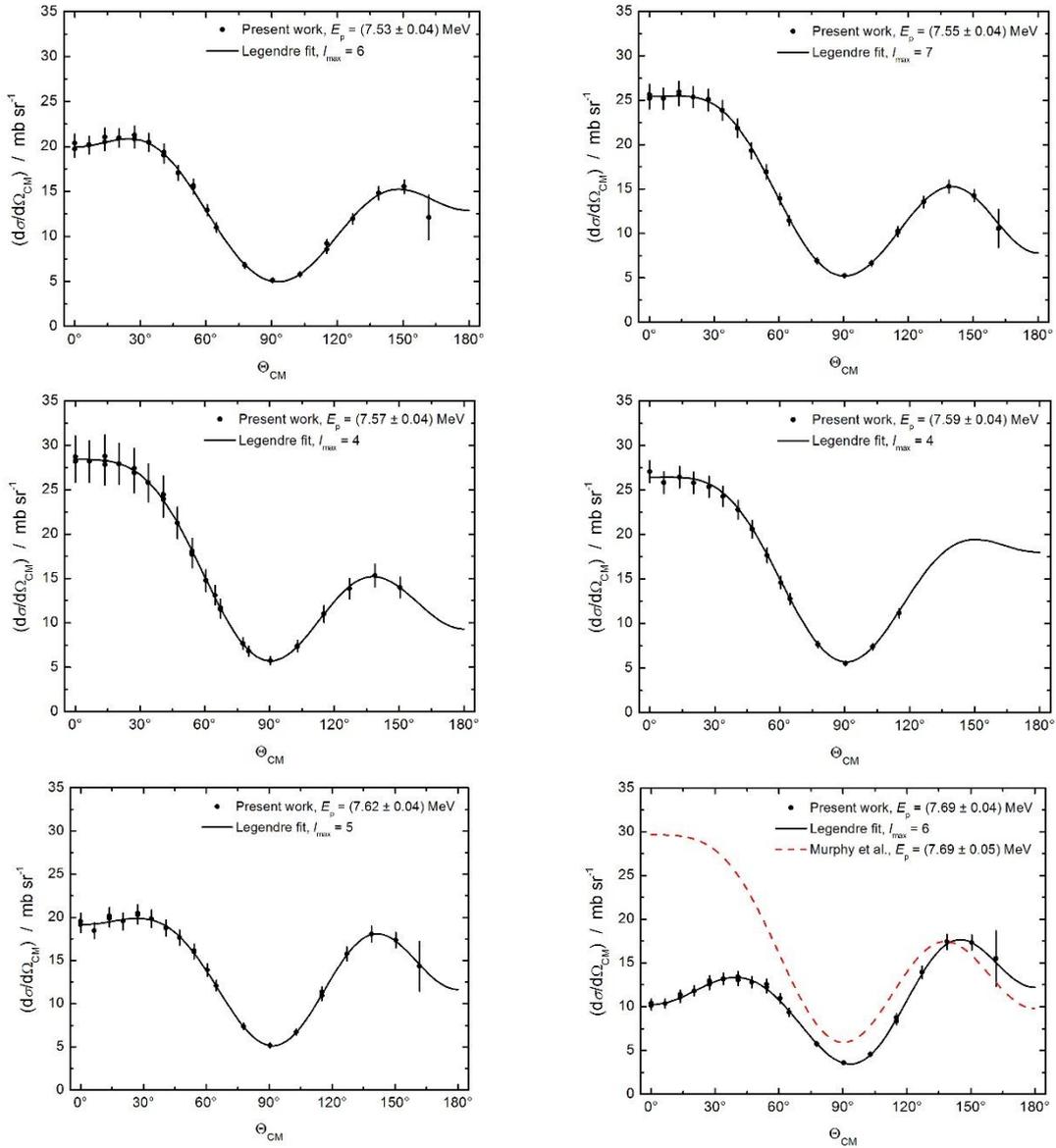

**Fig. 9** Differential cross sections of the $^{15}$N(p,n)$^{15}$O reaction for several proton energies around the resonance located at 7.59 MeV.

### 3.7 Differential 0° cross section from fluence measurements in scattering experiments

As mentioned earlier, measurements for the determination of scattering cross sections were successfully carried out using the $^{15}$N(p,n)$^{15}$O reaction as a neutron source [9, 10]. These experiments include measurements of the neutron fluence using detector D1 at an angle of 0° with respect to the deuteron beam line. These fluence measurements as well as measurements for testing purposes were used for the determination of differential cross sections for the $^{15}N(p,n)^{15}O$ reaction at 0°. The uncertainties are somewhat larger than for the measurement of complete angular distributions (7 % to 9 %) but are still smaller than for most of data in the literature. These results are included in the 0° excitation function for $^{15}N(p,n)^{15}O$ in addition to the differential cross section from dedicated measurements of the complete angular distributions (see figure 11 below).

The 0° fluence measurement at $E_p$ = 6.30 MeV ($E_n(0°)$ = 2.74 MeV) is of particular relevance because of the rather high 0° yield predicted by the DROSG-2000 code around this proton energy (see figure 11 below). Unfortunately, the angular distribution measured at $E_p$ = 6.32 MeV turned out to be compromised by problems with the proper extraction of the proton beam from the cyclotron. Hence, these data were excluded from the present work.

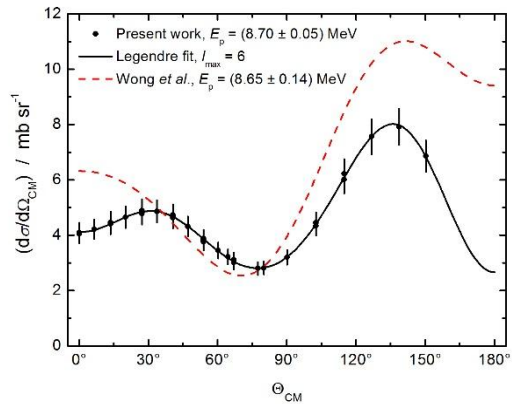

**Fig. 10** Differential cross sections of the $^{15}N(p,n)^{15}O$ reaction for a proton energy 8.70 MeV. The red dashed curve shows the angular distribution by Wong et al. for a proton energy 8.65 MeV.

### 3.8 Summary of the results

Differential 0° cross sections of this work are depicted in figure 11 together with data from the literature [12 - 19]. In contrast to figure 3, only those data are shown which were found in publications. The error bars for the energy do not represent the uncertainty of the proton energy (which is usually not stated) but the width of the proton energy distribution due to the energy loss in the target. The present data with complete angular distributions are indicated by the red closed circles. These data and their uncertainties were calculated from the Legendre polynomial expansions fitted to the experimental data. The uncertainties range from 2.8 % to 7.7 %. The results of the other fluence measurements at 0° are indicated by red open circles. The measurements of the present work confirm the large cross section at $E_p$ = 5.89 MeV ($E_n(0°)$ = 2.24 MeV) and $E_p$ = 6.33 MeV ($E_n(0°)$ = 2.74 MeV) in the 0° excitation function by Jones *et al.* [12] as well as for $E_p$ = 7.59 MeV ($E_n(0°)$ = 4.00 MeV) and $E_p$ = 8.9 MeV ($E_n(0°)$ = 5.3 MeV) which are marked with arrows in figure 3. The large 0° differential cross sections for $E_p$ = 6.65 MeV and $E_p$ = 7.7 MeV could not be confirmed.

Angle-integrated cross section for the same energy range are depicted in figure 12. The data of this work are shown by solid red circles. The uncertainties barely differ from the ones for the 0° differential cross section and range from 2.7 % to 7.7 %. All available data from references [12 - 19] are also shown in figure 12.

The data measured by Barnett [15] play a particularly important role. This data set includes more than 300 data points in the energy range $E_p$ = 3.746 MeV to 11.94 MeV. Because of the use of a very thin target and the resulting small energy width, resonances in the $^{15}$N(p,n)$^{15}$O cross section are easily visible. However, the cross sections are larger than the ones of most of the other measurements. In references [13, 14] it is concluded that the Barnett data (uncertainty 15 %) are 50 % to 60 % larger than the general trend. By linear regression, a linear parameterization of the cross section ratio was determined: $\sigma_{\text{Barnett}}/\sigma_{\text{PTB}} = 1.21 + 0.06 \cdot E_p / \text{MeV}$.

It results in adjustment factors of 0.59 to 0.67 in the energy range from $E_p$ = 5 MeV to 8 MeV which is in good agreement with references [13, 14]. The adjusted Barnett data are shown as a thick black curve in figure 12. As stated in reference [14], one can see that the simple adjustment of the Barnett data leads to a good agreement of most of the data sets. Part of the data by Hansen *et al.*, however, remain discrepant. Also, the data by Jones *et al.* for $E_p$ > 5.5 MeV deviate from the general trend with the deviation increasing with increasing proton energy.

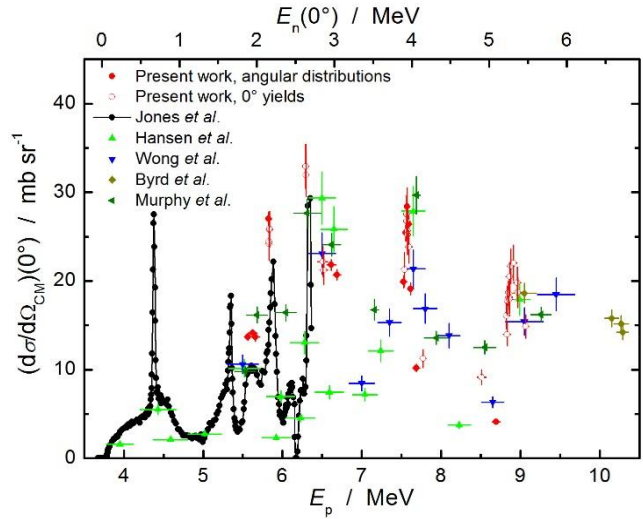

**Fig. 11** Differential cross sections for 0° of this work and data from the literature for comparison. Red closed circles show data sets with complete angular distributions. Red open circles show data measured only at $\Theta_{\text{LAB}}$ = 0°.

The cross section data measured by Chew *et al.* and Barnett for $E_p$ = 8.9 MeV to 9.4 MeV both exhibit a resonance structure with a similar width but a peak shifted by 100 keV. Perhaps some of the discrepancies were not caused by an uncertainty in the determination of the cross sections but by an uncertainty in the determination of the projectile energy.

The rescaling of the Barnett data for the angle-integrated cross section suggested by the present differential data has also important consequences for the production cross section of $^{15}$O via $^{15}$N(p,n)$^{15}$O. The radioisotope $^{15}$O is a β$^+$ emitter and plays an increasingly important role for positron emission tomography (PET). Besides the $^{14}$N(d,n)$^{15}$O reaction, the $^{15}$N(p,n)$^{15}$O reaction can be used for the production of $^{15}$O. In the energy range of this work, neutron production is only possible via the ground-state transition. Thus, the neutron production cross section is equal to the one for the production of $^{15}$O. In an evaluation of cross sections of the $^{15}$N(p,n)$^{15}$O reaction, Takács *et al.* [24] followed essentially the cross section data by Sajjad *et al.* [19], i.e. the Barnett data adjusted to fit these data. The cross sections of this work are approximately 20 % larger. The saturation activity, which is equivalent to the thick target yield $Y = \int (\Sigma(E)/(S(E)) \, dE$, is therefore 20 % larger if the Barnett data adjusted to the data of the present work are used. Here $\Sigma$ and $S$ denote the macroscopic cross section and the stopping power, respectively.

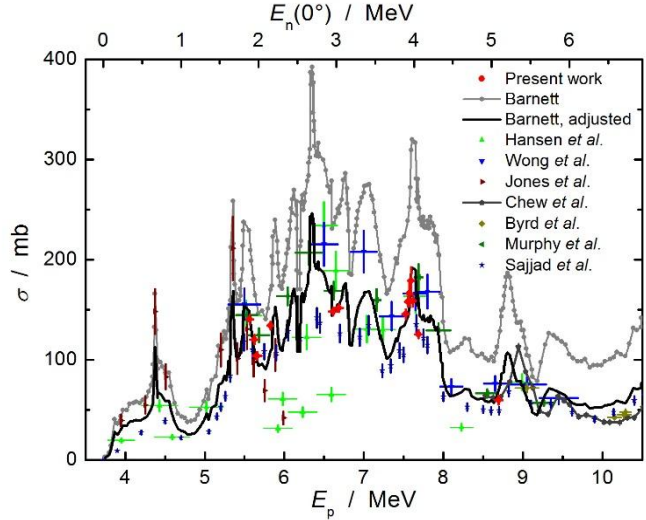

**Fig. 12** Cross sections of the $^{15}$N(p,n)$^{15}$O reaction of this work and data taken from the literature for comparison.

## 4. Discussion and conclusions

The $^{15}$N(p,n)$^{15}$O reaction was studied with an emphasis on the utilization as a source of quasi-monoenergetic neutrons. The aim is the extension of the neutron energy range of the PTB TOF spectrometer for the measurement of elastic and inelastic scattering cross sections to the low MeV region.

The $^{15}$N(p,n) reaction has several advantages. The reaction allows the production of quasi-monoenergetic neutrons for $E_n \leq 5.7$ MeV. $^{15}$N is a stable isotope and also the produced $^{15}$O activity does not pose a radiological hazard due to its short half life time of 122 s. With the $^{15}$N(p,n)$^{15}$O reaction, the advantages of the gas target can be used, i.e. easy variation of the target thickness by changing the gas pressure and easy background subtraction by measurement with a filled and an evacuated gas target cell. Moreover, highly enriched $^{15}$N is commercially available at moderate cost.

The reaction, however, also has several disadvantages. Because of the rather low 0° differential cross section and the large stopping power of $N_2$ gas, the neutron yield in the forward direction is approximately one order of magnitude smaller than for the D(d,n)$^3$He reaction. The excitation function has many small narrow resonances. Therefore, this reaction can be used as a neutron source for measurements of differential cross section experiments only at selected energies. Moreover, the strong variation of the angular distribution over the energy range (see figure 9) of the rather narrow resonances constitutes a potential source of problems when the energy of the proton beam is not well defined or unstable in time. This is a challenge for cyclotrons, in particular when magnetic analysis of the proton beam energy is missing which is the case for measurements at the PTB cyclotron.

For use as a neutron source, data for the angular distribution are required because the neutron energy distribution at forward angles will include neutrons produced at larger emission angles and scattered into the forward direction by material in the surroundings of the target. Usually, this contribution of scattered neutrons is assessed by Monte Carlo simulations which require differential cross section data as input. In this respect, the limited and sometimes discrepant cross section data for the $^{15}$N(p,n)$^{15}$O reaction are problematic.

Therefore, the PTB TOF spectrometer was used for the measurement of complete angular distributions and angle-integrated cross sections for 13 energies in the range $E_p$ = 5.56 MeV to 8.70 MeV. The uncertainties are in the order of 3 % to 8 %. In addition, fluence measurements carried out during the measurement of differential scattering cross sections were used to determine differential cross sections for 0° for 29 energies in the range from $E_p$ = 5.83 MeV to 9.07 MeV with uncertainties that range from 7 % to 9 %. The measurements do not cover all resonances in this energy range but lead to an improvement of the available cross section data and solve some discrepancies. For example, the large 0° differential cross section measured by Jones *et al.* for $E_p \approx 5.8$ MeV could be confirmed.

## Acknowledgments

The authors want to thank the staff of the PTB accelerator facility for providing excellent proton beams which met the challenging requirements of the present work. They are also grateful to H. Klein for his interest and support. This work was partly supported by the EFNUDAT project (contract number FP6-036434) funded by the European Commission.

## References

[1] D. Schmidt et al., *Precise time-of-flight spectrometry of fast neutrons – principles methods and results*, PTB report PTB-N-35, Braunschweig 1998, ISBN 3-89701-237-5006.

[2] D. Schmidt et al., *Determination of neutron scattering cross sections with high precision at PTB in the energy range 8 to 14 MeV*, Nucl. Sci. Eng. **160** (2008) 349 – 362

[3] G. Aliberti et al., *Impact of nuclear data uncertainties on transmuation of actinides in accelerator-driven assemblies*, Nucl. Sci. Eng **146** (2014) 13 – 50

[4] M. Embid et al. *Systematic uncertainties on Monte Carlo simulation of lead based ADS, actinide and fission product partitioning and transmutation*, Proceedings of the Fifth Information Exchange Meeting, Mol, Belgium, 25-27 November 1998

[5] M.Drosg, *DROSG-2000, codes and databases for 59 neutron source reactions*, IAEA report IAEA-NDS-87 Rev. 8, Vienna 2003

[6] J. Ziegler et al., *SRIM - The Stopping and Range of Ions in Matter*, Version 2003.26, retrieved online from www.srim.org

[7] D. Schmidt et al., *Differential Cross Sections of Neutron Scattering on Elemental Lead at Energies between 8 MeV and 14 MeV*, Report PTB-N-27, Braunschweig 1996, ISBN 3-89429-802-2

[8] G. Börker, *Messung des differentiellen Wirkungsquerschnittes der elastischen Streuung von Neutronen an Sauerstoff im Energiebereich von 6 MeV bis 15 MeV*, Dissertation, Ruhr-Universität Bochum 1987

[9] E. Poenitz et al., *Measurement of scattering cross sections of $^{nat}Pb$ at an incident neutron energy of 2.94 MeV*, Proceedings of the "International Conference on Nuclear Data for Science and Technology" (ND2007), Nizza, France, 22. - 27. April 2007; editors O. Bersillon, F. Gunsing, E. Bauge, R. Jacqmin, and S. Leray, EDP Sciences, 2008, 513 – 516 http://dx.doi.org/10.1051/ndata:07390

[10] E. Poenitz et al., *Elastic and inelastic neutron scattering cross section for $^{nat}Pb$, $^{209}Bi$ and $^{nat}Ta$ in the energy range from 2 MeV to 4 MeV*, J. Korean Phys. Soc. **59** (2011) 1876 - 1879

[11] D. Schmidt, B.R.L. Siebert, *Fast neutron spectrometry and Monte Carlo simulation – the codes SINENA and STREUER*, PTB report PTB-N-40, Braunschweig 2000, ISBN 3-89701-531-5

[12] K.W. Jones et al., *$^{15}N(p,n)^{15}O$ reaction study*, Phys. Rev. **112** (1958) 1252 – 1256

[13] S.H. Chew et al., *Resonance Structure in $^{15}N(p,n)^{15}O$ in the Region $E_p$ = 8.5 - 19.0 MeV*, Nucl. Phys. **A298** (1978) 19 – 30

[14] K. Murphy et al., The *$^{15}N(p,n)^{15}O$ reaction below 9.3 MeV*, Nucl. Phys. **A355** (1981) 1 – 12

[15] A.R.Barnett, *$^{16}O$ analogue states in the $^{15}N(p,n)^{15}O$ reaction*, Nucl. Phys. **A120** (1968) 342 – 368

[16] L.F. Hansen, M.L. Stelts, *$^{15}N(p,n)^{15}O$ ground-state reactions and the quasielastic model of (p,n) reactions*, Phys. Rev. **132** (1963) 1123 – 1130

[17] C. Wong et al., *Angular distribution of the ground-state neutrons from the $^{13}C(p,n)^{13}N$ and $^{15}N(p,n)^{15}O$ reactions*, Phys. Rev. **123** (1961) 598 – 605

[18] R.C.Byrd et al., *Measurement and Lane-model analysis of cross sections for the $^{13}C(p,n)^{13}N$ and $^{15}N(p,n)^{15}O$ reactions*, Nucl. Phys. **A351** (1981) 189 – 218

[19] M. Sajjad et al., *Excitation function for the $^{15}N(p,n)^{15}O$ reaction*, Journal of Labelled Compounds and Radiopharmaceuticals **21** (1984) 1260

[20] L.P. Robertson et al., *Beam heating effects in gas targets*, Rev. Sci. Instr. **32** (1961) 1405

[21] D. Schmidt et al., *Investigation of the $^9Be(\alpha,n)^{12}C$ reaction, I: Experimental procedure and uncertainties*, PTB Report PTB-N-7, Braunschweig 1992, ISBN 3-89429-176-1

[22] G. Dietze et al., *NRESP4 and NEFF4 – Monte Carlo codes for the calculation of neutron response functions and detection efficiencies for NE213 scintillation detectors*, PTB Report PTB-ND-22, Braunschweig 1982, ISSN 0572-7170

[23] E. Pönitz et al., EXFOR entries 02099001 - 02099005, http://www-nds.iaea.org/EXFOR/23156001

[24] S. Takács et al., *Validation and upgrading of the recommended cross section data of charged particle reactions used for production of PET radioisotopes*, Nucl. Instr. and Methods **B211** (2003) 169 – 189